\documentclass[11pt,twoside]{article}
\usepackage{asp2010}
\usepackage{graphicx}
\resetcounters

\begin{document}

\title{Distribution of High Mass X-ray Binaries in the Milky Way}
\author{Alexis Coleiro and Sylvain Chaty
\affil{Laboratoire AIM (UMR 7158 CEA/DSM - CNRS - Universit\'e Paris Diderot), Irfu / Service d'Astrophysique, CEA-Saclay, 91191 Gif-sur-Yvette Cedex, France. e-mail: alexis.coleiro@cea.fr; chaty@cea.fr}}

\begin{abstract}
The INTEGRAL satellite, observing the sky at high energy, has quadrupled the number of supergiant X-ray Binaries known in the Galaxy and has revealed new populations of previously hidden High Mass X-ray Binaries. These observations raise new questions about the formation and evolution of these sources. The number of detected sources is now high enough to allow us to carry out a statistical analysis of the distribution of HMXBs in the Milky Way.
We derive the distance of each HMXB using a Spectral Energy Distribution fitting procedure, and we examine the correlation with the distribution of star forming complexes (SFCs) in the Galaxy. We show that HMXBs are clustered with SFCs, with a typical size of 0.3 kpc and a characteristic distance between clusters of 1.7 kpc. 
\end{abstract}

\section{Introduction}

High Mass X-ray Binaries (HMXBs) are binary systems composed of a compact object, a neutron star or a black hole candidate, accreting matter from a massive companion star: either a main sequence Be star or an evolved supergiant O or B star. Most of these sources are observed in the Galactic Plane \citep{Bird_2007} as it is expected for such young star systems which do not have time to move far from their birthplaces.\\
Thanks to the dedicated observations from \textit{RXTE} and \textit{INTEGRAL}, around 200 HMXBs are currently known in the Milky Way allowing us to focus on their distribution. Using RXTE data, \citet{Grimm_2002} highlighted clear signatures of the spiral structure in the spatial distribution of HMXBs. In the same way, \citet{Dean_2005}, \citet{Lutovinov_2005}, \citet{Bodaghee_2007} and \citet{Bodaghee_2011} showed that HMXBs observed with \textit{INTEGRAL} also seem to be associated with the spiral structure of the Galaxy. However, the HMXB positions, mostly derived from their X-ray luminosity, are not well constrained and highly uncertain due to direct accretion as in HMXB. In order to overcome this caveat we present a novel approach allowing us to derive all HMXB positions.
We build the Spectral Energy Distribution (SED) of each HMXB and fit it with a black-body model to compute the distance of each source.
Finally, we study this distribution and the correlation with Star Forming Complexes (SFCs) observed in the Galaxy. 

\section{Deriving the HMXB location within the Galaxy}

In order to compute the HMXB distances, we gathered a set of HMXBs for which at least 4 optical and/or NIR magnitudes are known, and for each source its SED was built and fit with a blackbody model. This enables us to evaluate the distance of the source along with its associated errors.

\subsection{Whole catalogue of HMXBs}

There are currently more than 200 HMXBs detected in the Milky Way. Using an updated version of the \citet{Liu_2006} catalogue\footnote{see IGR Sources web page maintained by J. Rodriguez \& A. Bodaghee (http://irfu.cea.fr/Sap/IGR-Sources/)}, we retrieved the optical, NIR magnitudes, the spectral type and the luminosity class of each source. For each HMXB, four magnitudes are required to compute the fitting procedure. Indeed, according to the $\chi^2$ computation, the condition $N-n \ge 2$ with $N$ the number of observed magnitudes and $n$ the number of parameters left free (in our study, $n=2$, cf. section \ref{fit}), needs to be met. Finally, we selected around 70 sources meeting these conditions.

\subsection{SED fitting procedure}\label{fit}

\begin{figure}[h]
\begin{center}
\includegraphics[scale=0.2]{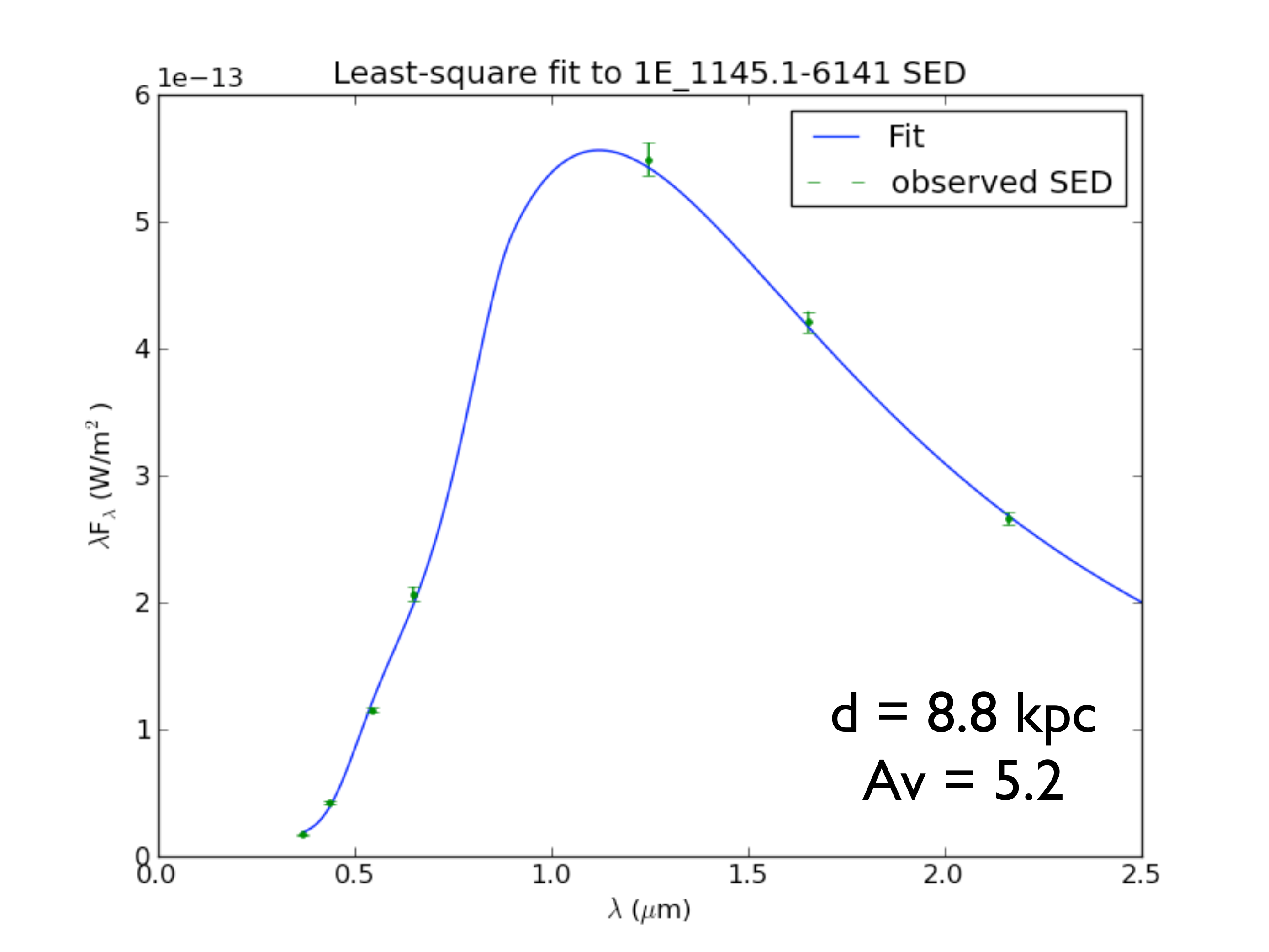}
\caption{Result of the fitting for one source with distance $D$ and extinction in V band, $A_{\rm{V}}$.}
\label{fits}
\end{center}
\end{figure}

Our fitting procedure is based on the Levenberg-Marquardt least-square algorithm implemented in Python.  For each HMXB, we build the SED in optical and NIR from a maximum of 8 magnitude points (U, B, V, R, I, J, H, Ks) and, as we underlined before, a minimum of 4 magnitudes are required by the fitting. This SED is then fit (see figure \ref{fits}) by a black-body model given by the relation:

\begin{equation}
\lambda F_{\lambda} =\frac{2\pi hc^2}{\lambda^4}\times10^{-0.4A_{\lambda}}\frac{\left(R/D\right)^2}{\exp{\left(\frac{hc}{\lambda k_{B}T}\right)}}
\label{blackbody}
\end{equation}

where $\lambda$ is the wavelength in $\mu$m, $F_{\lambda}$ the flux density in W.m$^{-2}$.$\mu$m$^{-1}$, $h$ the Planck constant, $k_B$ the Botzmann constant, $c$ the speed of light, $A_\lambda$ the extinction at the wavelength $\lambda$, $R/D$ the stellar radius over distance ratio and $T$ the temperature of the star. From the spectral type and the luminosity class, we derive the radius and the temperature of the companion star which dominates the optical and NIR flux \citep{Morton_1968, Panagia_1973, Searle_2008}. Two parameters are left free: the extinction in V band $A_{\rm{V}}$ and the ratio $R/D$ whereas the extinction $A_\lambda$ is derived from \citet{Cardelli_1989} at each wavelength assuming $R_{\rm{V}}=3.1$ for the Milky Way. Knowing the radius $R$ of the companion star (see \citealt{Morton_1968}, \citealt{Panagia_1973}) , we calculate the distance $D$ in kpc.\\

The least square function given by the formula 

\begin{equation}
\chi^2=\sum_i\left[\frac{X_{i,obs}-X_{i,model}}{\sigma_i}\right]^2
\label{lstquare}
\end{equation}

(with $X_{i,obs}$ the observed flux value for the filter $i$, $X_{i,model}$, the theoretical flux in the i$^{th}$ filter derived from the black-body model and $\sigma_i$, the flux error in the same filter) is then minimized by the Levenberg-Marquardt algorithm.

\subsection{Uncertainties in the computed distances}
The magnitude uncertainties are retrieved from the literature. For the sources for which no error is given, we use a systematic error of 0.1 magnitude. The flux uncertainties are then derived from these magnitude errors. Otherwise, we assume the spectral type given in the literature as the real spectral type of the companion star. Simulations were carried out and enabled us to constrain the spectral type error that appears to be less important for supergiant stars than for Zero-Age Main Sequence ones.

Degeneracy between several parameters values (based on the fitting procedure) needs to be taken into account. Indeed, solely relying on a single best fit does not capture the full phenomenology associated with SED fitting because extinction $A_{\rm{V}}$ and distance $D$ are degenerated in this approach. In order to produce the most likely set of fits and to determine the dispersion on distance and extinction we carried out 500 Monte-Carlo simulations for each observed source by varying the photometry within the uncertainties. Hence, we generate a random number from a normal distribution (assuming the photometric errors to be Gaussian) contained within the error bars for each photometric point so that we build 500 new SEDs derived from the original one. These 500 new SEDs are subject to the same $\chi^2$ statistic computation as the one described above. Then, we have an entire set of best fits of parameters ($A_{\rm{V}}$, $D$) and, we are able to plot the distribution in the parameters space, showing the distribution of properties derived from these Monte-Carlo simulations and especially showing the dispersion on distance and extinction. This dispersion value is taken as the error coming from the fitting procedure and the median value of dispersion on distance determination (for all the HMXBs under study) is 0.75 kpc.\\

There are other sources of uncertainties, particularly the infrared excess of Be stars due to their circumstellar envelope generating free-free radiation. According to \citet{Dougherty_1994}, this excess should not exceed a mean 0.1 magnitude in J band, 0.15 magnitude in H band and 0.25 magnitude in K band. However, this value corresponds to absolute magnitude and therefore can be smaller for sources located far away and higher for close ones. To take this effect into account, an estimate of the distance and extinction $A_{\rm{V}}$ is needed. Since these two values are derived from the fitting procedure, we are only able to consider as a distance and absorption estimate the values obtained without taking this IR excess into account. This approach is finally equivalent to adding a conservative excess of 0.1, 0.15 and 0.25 magnitude to the apparent magnitude in the J, H and Ks bands respectively. The results are presented on figure \ref{unc}.1.

\articlefiguretwo{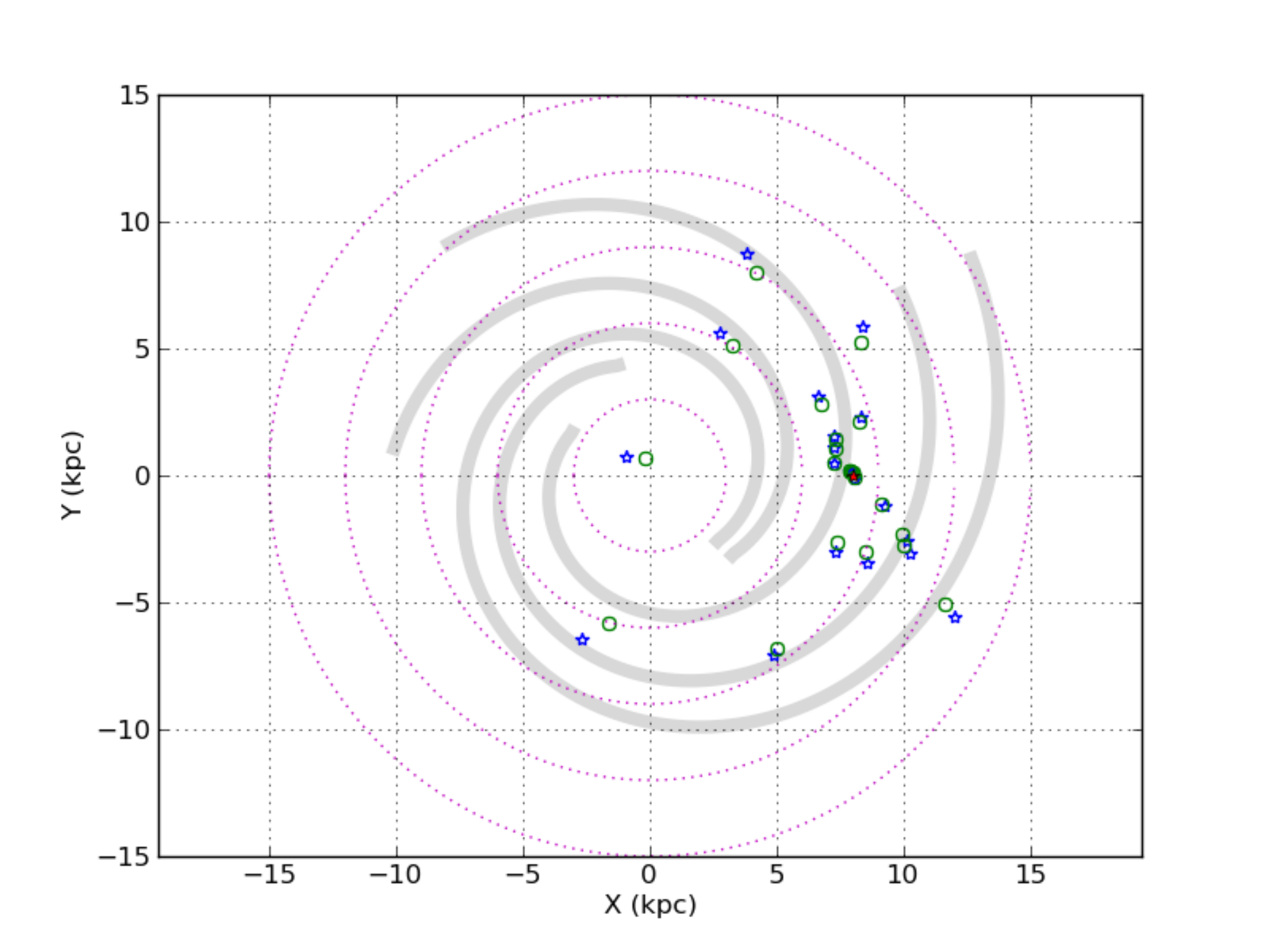}{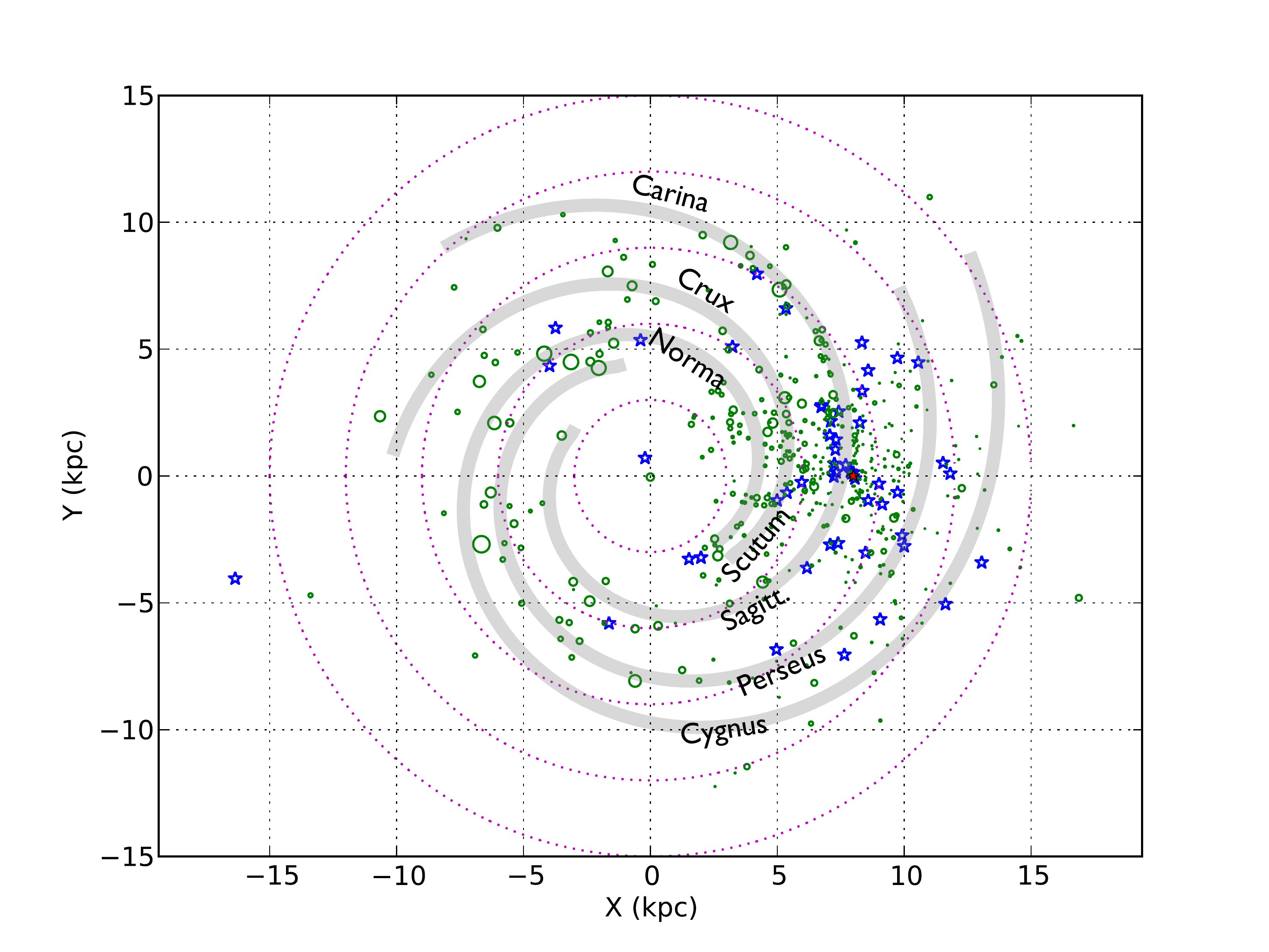}{unc}{1) Green circles represent the initial positions of Be stars whereas blue stars represent the source positions taking into account the IR excess. 2) Distribution of HMXBs (blue stars) and distribution of SFCs (green circles). Circle radius represent the different excitation parameter values. The spiral arms model from \citet{Russeil_2003} is also plotted and the red star (at X = 8 kpc, Y = 0 kpc) represents the Sun position at 8 kpc from the Galactic center.}

\section{Results: distribution of HMXBs and correlation with Star Forming Complexes distribution}

We present hereafter the distribution of HMXBs in the Galaxy, obtained with our novel approach. The spiral arms model given by \citet{Russeil_2003} is also presented (see figure \ref{unc}.2). We want to assess the question whether there is a correlation between this distribution of HMXBs and, in a first time, the distribution of Star Forming Complexes (SFCs) in the Milky Way (given by \citealt{Russeil_2003}), as it is expected from the short HMXB lifetime.\\

 The first approach we adopt is to carry out a Kolmogorov-Smirnov test (KS-test) in each axis in order to quantify the fact that the two samples are drawn or not from the same probability distribution. We got a value of 0.15 for the X axis, a value of 0.25 for the Y axis and a value of 0.31 for the galactic longitude. Even if these values are not negligible, suggesting that a correlation between the two samples is likely established, part of the information is lost because of the projection on the two axis. To overcome this caveat, we propose another method described hereafter.\\

We suppose that each HMXB (blue stars on figure \ref{correl1}.1) is clustered with several SFCs (green circles on the same figure). Hence, we can define two characteristic scales: a typical cluster size and a typical distance between clusters. Around each HMXB, we define several circles with different radius (red circles on figure \ref{correl1}.1) and we finally count the number of HMXBs for which at least one SFC is within the specified radius.

\articlefiguretwo{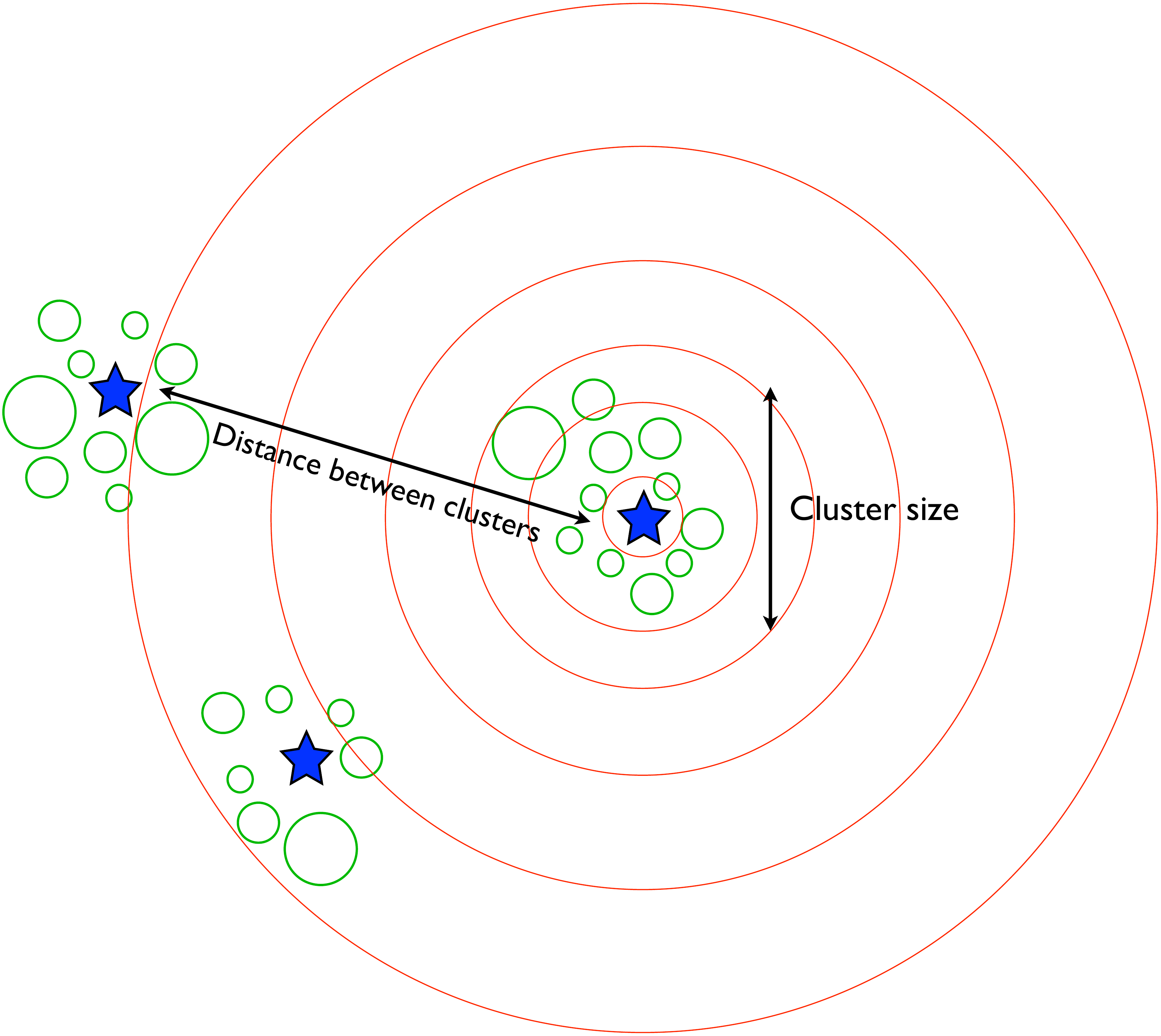}{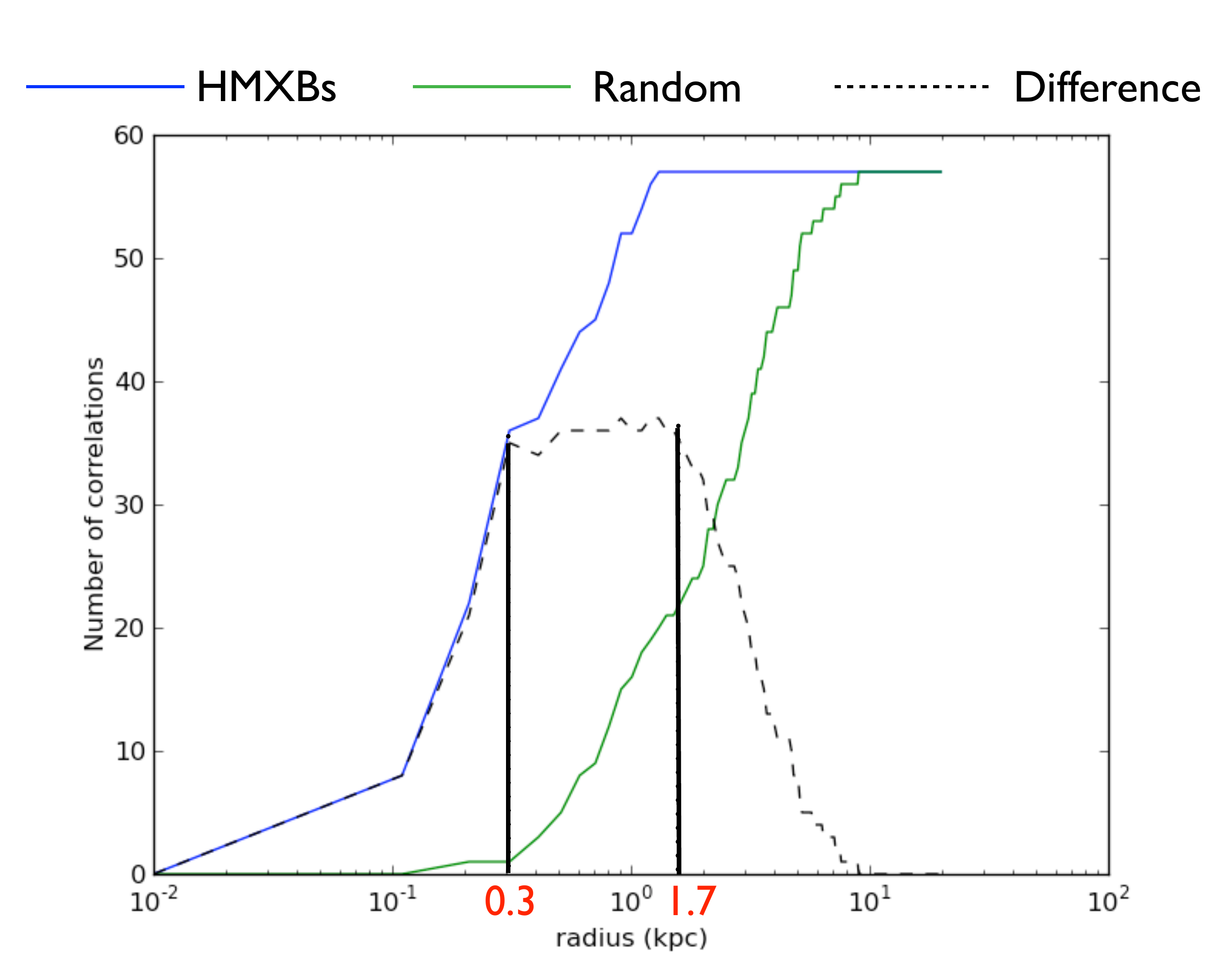}{correl1}{ 1) Description of the method used to evaluate the correlation. 2) Result of the correlation determination in 2D.}

The number of HMXBs for which at least one SFC is within the specified radius versus the circle radius is plotted as the blue curve on figure \ref{correl1}.2. The green curve is that expected from chance correlations assuming the HMXBs are evenly distributed across the sky. The dashed line represents the difference between the two previous curves. This difference, being non equal to zero, allows us to state that a strong correlation exists between the HMXB and the SFC positions in the Milky Way. Moreover, we compute the two characteristic scales exposed before: the typical cluster size of 0.3 kpc and the typical distance between clusters of 1.7 kpc, which is larger than the median error on HMXB distances.
If we take into account the uncertainties in the HMXB positions (obtained via the Monte-Carlo simulations) and in the SFC positions (given in \citealt{Russeil_2003}, median error of 0.25 kpc), the correlation still exists with the same cluster size and the same distance between clusters. These results, while obtained with a different method, are consistent with those reached by \citet{Bodaghee_2011}. Finally, we test our correlation code using a sample of globular clusters (\citealt{Bica_2006}), principally located in the Galactic bulge. Therefore, figure \ref{GC}.1 shows the two distributions in the Milky Way whereas figure \ref{GC}.2 shows the result of correlation test. Clearly, as expected, no correlation is revealed.

\articlefiguretwo{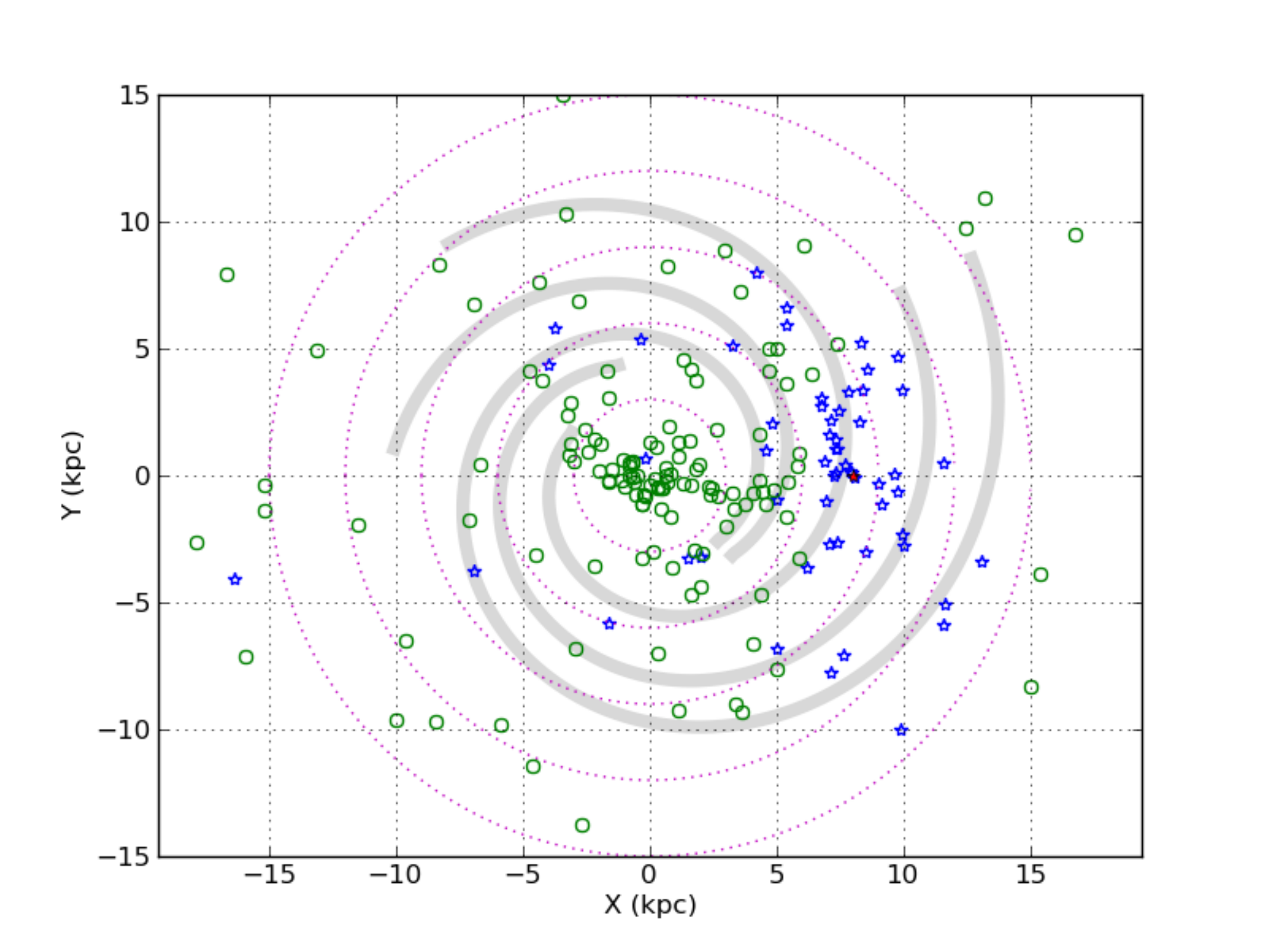}{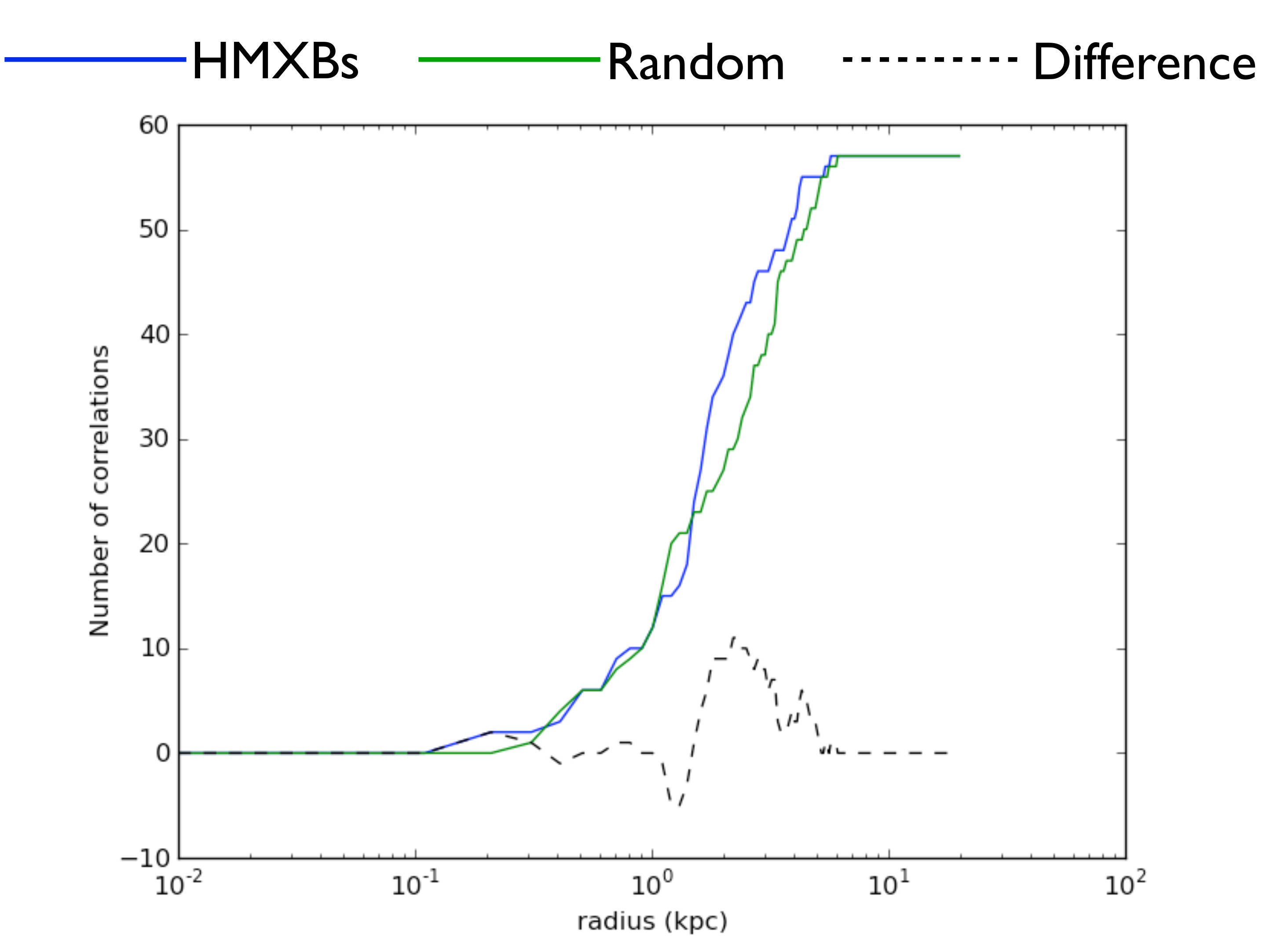}{GC}{1) Distribution of HMXBs and globular clusters. 2) Result of the correlation determination with globular clusters.}

\section{Conclusion}
Evaluation of HMXB distribution is now of major interest in order to study in depth the formation of these high energy sources. However, HMXB locations are usually poorly constrained and largely dependent on the determination method. We propose here to determine the location of a whole set of sources using the same approach: a SED fitting of the distance of HMXBs. This method, based on a least-square minimization, enables us to reveal a consistent picture of the HMXB distribution, showing them to follow the spiral arm structure of the Galaxy. The consideration of uncertainties leads to a small error on the source locations and allows us to tackle the study of the correlation with SFC distribution. This study shows that HMXBs are clustered with SFCs and enables for the first time to quantitatively define the cluster size (0.3 kpc) and the distance between clusters (1.7 kpc). The challenge is now to quantify this correlation by taking into account the offset between current spiral density wave position and HMXB positions expected due to the fact that the matter rotation velocity is different to the spiral arm rotation speed.

\acknowledgments  We acknowledge A. Bodaghee for his help and discussions about this project. We acknowledge P.A. Charles, C. Knigge, J. A. Zurita Heras, P. A. Curran and F. Rahoui for useful discussions and advice. This work was supported by the Centre National d'Etudes Spatiales (CNES), based on observations obtained with MINE -- the Multi-wavelength INTEGRAL NEtwork --. This research has made use of the IGR Sources page maintained by J. Rodriguez \& A. Bodaghee (http://irfu.cea.fr/Sap/IGR-Sources/), of data products from the Two Micron All Sky Survey, which is a joint project of the University of Massachusetts and the Infrared Processing and Analysis Center/California Institute of Technology, funded by the National Aeronautics and Space Administration and the National Science Foundation and of the SIMBAD database and the VizieR catalogue access tool, operated at CDS, Strasbourg, France.

\bibliography{biblio_1}

\begin{thebibliography}{}
\expandafter\ifx\csname natexlab\endcsname\relax\def\natexlab#1{#1}\fi
\expandafter\ifx\csname url\endcsname\relax
  \def\url#1{\texttt{#1}}\fi
\expandafter\ifx\csname urlprefix\endcsname\relax\def\urlprefix{URL }\fi
\providecommand{\eprint}[2][]{\url{#2}}

\bibitem[{{Bica} et~al.(2006){Bica}, {Bonatto}, {Barbuy}, \&
  {Ortolani}}]{Bica_2006}
{Bica}, E., {Bonatto}, C., {Barbuy}, B., \& {Ortolani}, S. 2006, \aap, 450, 105

\bibitem[{{Bird} et~al.(2007){Bird}, {Malizia}, {Bazzano}, {Barlow}, {Bassani},
  {Hill}, {B{\'e}langer}, {Capitanio}, {Clark}, {Dean}, {Fiocchi}, {G{\"o}tz},
  {Lebrun}, {Molina}, {Produit}, {Renaud}, {Sguera}, {Stephen}, {Terrier},
  {Ubertini}, {Walter}, {Winkler}, \& {Zurita}}]{Bird_2007}
{Bird}, A.~J., {Malizia}, A., {Bazzano}, A., {Barlow}, E.~J., {Bassani}, L.,
  {Hill}, A.~B., {B{\'e}langer}, G., {Capitanio}, F., {Clark}, D.~J., {Dean},
  A.~J., {Fiocchi}, M., {G{\"o}tz}, D., {Lebrun}, F., {Molina}, M., {Produit},
  N., {Renaud}, M., {Sguera}, V., {Stephen}, J.~B., {Terrier}, R., {Ubertini},
  P., {Walter}, R., {Winkler}, C., \& {Zurita}, J. 2007, \apjs, 170, 175

\bibitem[{{Bodaghee} et~al.(2007){Bodaghee}, {Courvoisier}, {Rodriguez},
  {Beckmann}, {Produit}, {Hannikainen}, {Kuulkers}, {Willis}, \&
  {Wendt}}]{Bodaghee_2007}
{Bodaghee}, A., {Courvoisier}, T., {Rodriguez}, J., {Beckmann}, V., {Produit},
  N., {Hannikainen}, D., {Kuulkers}, E., {Willis}, D.~R., \& {Wendt}, G. 2007,
  \aap, 467, 585

\bibitem[{{Bodaghee} et~al.(2011){Bodaghee}, {Tomsick}, \&
  {Rodriguez}}]{Bodaghee_2011}
{Bodaghee}, A., {Tomsick}, J.~A., \& {Rodriguez}, J. 2011, ArXiv e-prints.
  \eprint{1102.3666}

\bibitem[{{Cardelli} et~al.(1989){Cardelli}, {Clayton}, \&
  {Mathis}}]{Cardelli_1989}
{Cardelli}, J.~A., {Clayton}, G.~C., \& {Mathis}, J.~S. 1989, \apj, 345, 245

\bibitem[{{Dean} et~al.(2005){Dean}, {Bazzano}, {Hill}, {Stephen}, {Bassani},
  {Barlow}, {Bird}, {Lebrun}, {Sguera}, {Shaw}, {Ubertini}, {Walter}, \&
  {Willis}}]{Dean_2005}
{Dean}, A.~J., {Bazzano}, A., {Hill}, A.~B., {Stephen}, J.~B., {Bassani}, L.,
  {Barlow}, E.~J., {Bird}, A.~J., {Lebrun}, F., {Sguera}, V., {Shaw}, S.~E.,
  {Ubertini}, P., {Walter}, R., \& {Willis}, D.~R. 2005, \aap, 443, 485.
  \eprint{arXiv:astro-ph/0508291}

\bibitem[{{Dougherty} et~al.(1994){Dougherty}, {Waters}, {Burki}, {Cote},
  {Cramer}, {van Kerkwijk}, \& {Taylor}}]{Dougherty_1994}
{Dougherty}, S.~M., {Waters}, L.~B.~F.~M., {Burki}, G., {Cote}, J., {Cramer},
  N., {van Kerkwijk}, M.~H., \& {Taylor}, A.~R. 1994, \aap, 290, 609

\bibitem[{{Grimm} et~al.(2002){Grimm}, {Gilfanov}, \& {Sunyaev}}]{Grimm_2002}
{Grimm}, H., {Gilfanov}, M., \& {Sunyaev}, R. 2002, \aap, 391, 923.
  \eprint{arXiv:astro-ph/0109239}

\bibitem[{{Liu} et~al.(2006){Liu}, {van Paradijs}, \& {van den
  Heuvel}}]{Liu_2006}
{Liu}, Q.~Z., {van Paradijs}, J., \& {van den Heuvel}, E.~P.~J. 2006, \aap,
  455, 1165. \eprint{0707.0549}

\bibitem[{{Lutovinov} et~al.(2005){Lutovinov}, {Revnivtsev}, {Gilfanov},
  {Shtykovskiy}, {Molkov}, \& {Sunyaev}}]{Lutovinov_2005}
{Lutovinov}, A., {Revnivtsev}, M., {Gilfanov}, M., {Shtykovskiy}, P., {Molkov},
  S., \& {Sunyaev}, R. 2005, \aap, 444, 821

\bibitem[{{Morton} \& {Adams}(1968)}]{Morton_1968}
{Morton}, D.~C., \& {Adams}, T.~F. 1968, \apj, 151, 611

\bibitem[{{Panagia}(1973)}]{Panagia_1973}
{Panagia}, N. 1973, \aj, 78, 929

\bibitem[{{Russeil}(2003)}]{Russeil_2003}
{Russeil}, D. 2003, \aap, 397, 133

\bibitem[{{Searle} et~al.(2008){Searle}, {Prinja}, {Massa}, \&
  {Ryans}}]{Searle_2008}
{Searle}, S.~C., {Prinja}, R.~K., {Massa}, D., \& {Ryans}, R. 2008, \aap, 481,
  777

\end{thebibliography}

\end{document}